\documentclass[
reprint,
longbibliography,
bibnotes,
amsmath,amssymb,
aps,
prb,
floatfix
]{revtex4-2}
\usepackage{hyperref}
    \hypersetup{colorlinks, citecolor=black, linkcolor= black, urlcolor=blue}
\usepackage{graphicx}
\usepackage{dcolumn}
\usepackage{bm}
\usepackage{pgf}
\usepackage{physics}
\usepackage{dsfont}
\usepackage{xcolor}
\newcommand{\mvcm}{\frac{\text{MV}}{\text{cm}}}
\begin{document}
\preprint{APS/123-QED}
\title{Ultrafast Electrons in Noble Metals: Orientational Relaxation, Thermalization and Cooling in Terms of Electron-Phonon Interaction}
\author{Jonas Grumm}
\email{j.grumm@tu-berlin.de}
\author{Andreas Knorr}
\email{andreas.knorr@tu-berlin.de}
\affiliation{Nichtlineare Optik und Quantenelektronik, Institut für Physik und Astronomie, Technische Universität Berlin, 10623 Berlin, Germany}
\date{\today}
\begin{abstract}
We investigate the momentum-resolved dynamics of conduction electrons in noble metals following ultrashort optical excitation with linearly polarized light. Using a momentum-resolved Boltzmann equation approach for electron-phonon interaction, we solve for the combined effects of orientational relaxation, thermalization, and cooling. We introduce momentum orientational relaxation as the initial step in the equilibration of an optically excited non-equilibrium electron gas by highlighting its importance for the optical response of noble metals and the dephasing of plasmonic excitations. Our numerical results for gold reveal that orientational relaxation exists independently of the absorbed optical energy and dominates on time scales on the first tens of fs after excitation. Incorporating also thermalization and cooling on times up to a few ps, our approach provides a simultaneous description of optical and thermal properties of noble metals under initial non-equilibrium conditions.
\end{abstract}
\maketitle
\begin{figure}[b]
    \includegraphics[width=1\linewidth]{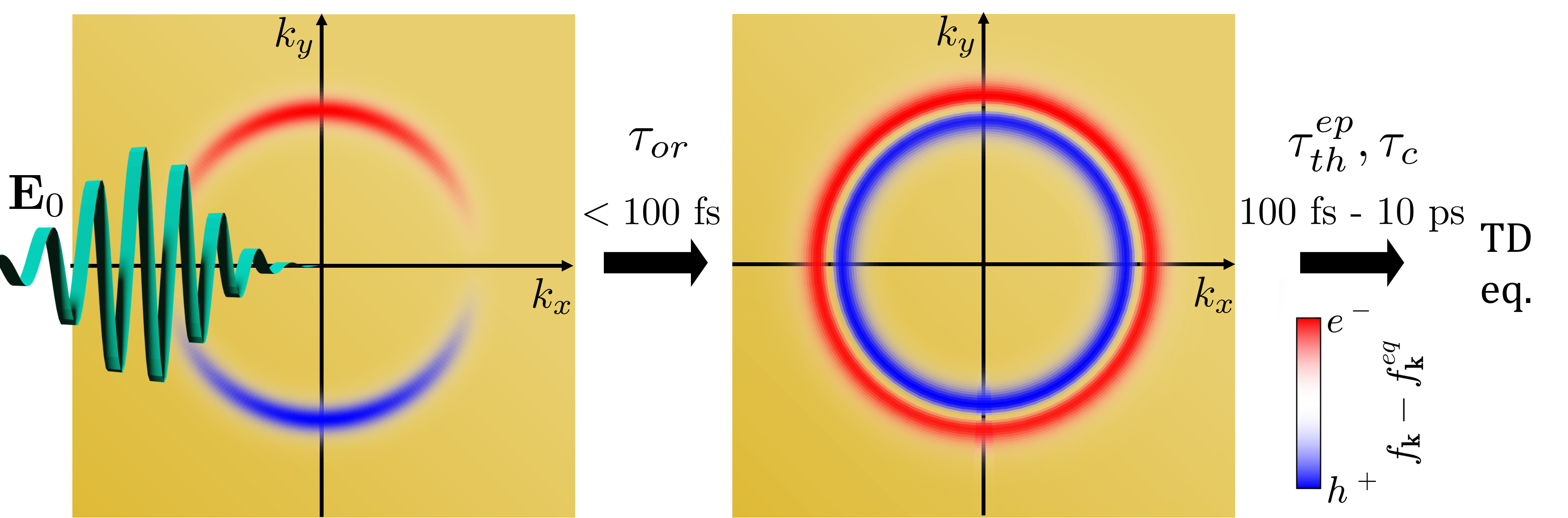} 
    \caption{\label{fig:ORT} A linear polarized light field $\vb{E}_0$ induces a momentum-polarization of the electron distribution $f_\mathbf{k}$ creating excited intraband electrons and holes near the Fermi edge. This state dominates on the first tens of fs after excitation. The excited non-symmetric distribution decays in the orientational relaxation time $\tau_{or}$ into a symmetric distribution. The relaxation back into the thermodynamic equilibrium (TD eq.) proceeds in the thermalization and cooling time ($\tau_{th}^{ep}$, $\tau_c$) on longer time scales up to several ps.}
\end{figure}

\textit{Introduction.~~} The ultrashort excitation of electrons in noble metals by light fields with high intensities results in extreme non-equilibrium conditions. In the recent years, the development of such ultrashort excitations made it possible to observe of the dynamical optical response of solids with femtosecond resolution and provides new insights into the built-up and decay of non-equilibrium electronic or plasmonic excitations \cite{krausz_attosecond_2014, calegari_advances_2016, bionta_-chip_2021, bionta_tracking_2021, zimin_dynamic_2023, wong_far-field_2024}. The relaxation back to a thermodynamic equilibrium, which proceeds in noble metals on scales from a few fs to several ps, depends strongly on the excitation conditions. 
The relaxation of the non-equilibrium energy-resolved electron distribution itself can be divided in thermalization by electron-electron interaction \cite{seibel_time-resolved_2023}, dissipation (electron cooling) by electron-phonon interaction \cite{del_fatti_nonequilibrium_2000, pietanza_non-equilibrium_2007, rethfeld_ultrafast_2002, mueller_relaxation_2013, ono_ultrafast_2020, riffe_excitation_2023} and lattice cooling by phonon-phonon interaction \cite{hartland_coherent_2002, salzwedel_theory_2023}. These processes are addressed in many publications by effective macroscopic models \cite{sun_femtosecond-tunable_1994, della_valle_real-time_2012, sivan_nonlinear_2017, ndione_nonequilibrium_2022, schirato_ultrafast_2023} or microscopic theories based on the semi-classical energy-resolved Boltzmann equation \cite{allen_theory_1987, dubi_hot_2019, sivan_theory_2021, del_fatti_nonequilibrium_2000, pietanza_non-equilibrium_2007, rethfeld_ultrafast_2002, mueller_relaxation_2013, seibel_time-resolved_2023, riffe_excitation_2023, ono_ultrafast_2020, sarkar_electronic_2023}.
\\%
These energy-resolved approaches neglect the fact that the polarization of the exciting light field breaks the symmetry of the initially isotropic electron gas. This symmetry break corresponds to an anisotropic excitation of electronic momentum which cannot be addressed in an energy-resolved basis only. However, the recent development of pulses short as several fs should make it possible to connect momentum-resolved measurements by using orthogonal polarized pump-probe experiments. 
\\%
In this Letter, we introduce the orientational relaxation of the momentum-polarized conduction electrons in noble metals as the first equilibrization step after optical excitation with linearly polarized light (cf.~Fig.~\ref{fig:ORT}). We show that the orientational relaxation is fundamental for the optical response of noble metals \cite{grumm_theory_2025}, acts as dephasing of the collectively excited electrons (plasmons) in the formation of hot electrons, and effects the energy relaxation dynamics up to $100$~fs after the excitation.
So far, to our knowledge, momentum orientational relaxation was never discussed beyond the relaxation time approximation \cite{ziman_electrons_1960, lawrence_umklapp_1972, brown_nonradiative_2016, brown_ab_2016, mustafa_ab_2016} but occurs on the same footing with thermalization and cooling of a non-equilibrium electron gas as a first dephasing process. To tackle this problem, we consider the numerical solution of a momentum-resolved Boltzmann equation exemplarily for gold including light-matter interaction and electron-phonon scattering \cite{grumm_theory_2025}. Our approach includes orientational relaxation, thermalization as redistribution of electronic energy and cooling as energy dissipation from electrons to phonons. To study a well defined system, all three processes are described by a consistent treatment of the electron-phonon interaction via normal and umklapp processes. In the following, we will discuss these three relaxation processes in terms of the dynamics of the electron distribution and introduce characteristic time scales via observables.

\textit{Theory.~~}
For the description of the dynamics of electrons in the \textit{sp}-conduction band of noble metals after optical excitation, we apply a parabolic and isotropic dispersion $\epsilon_k = \frac{1}{2m}\hbar^2 |\vb{k}|^2$ at their Fermi edge with an effective mass $m$ at room temperature $T_{eq}=300$~K. For optical excitations below the electronic band gap, a description via intraband excitations is sufficient \cite{brown_ab_2016, sundararaman_theoretical_2014, mustafa_ab_2016}. Dissipative processes are incorporated via a screened electron-phonon coupling \cite{mahan_solid_2000, ashcroft_solid_1976, del_fatti_nonequilibrium_2000, mueller_relaxation_2013}. 
By introducing the microscopic electronic conduction band occupations $f_\mathbf{k}$ with momentum $\vb{k}$, we obtain \cite{grumm_theory_2025} the momentum-resolved Boltzmann equation \cite{riffe_excitation_2023, salzwedel_theory_2023, meier_coherent_1994}
\begin{align}
    \partial_t f_\mathbf{k}(t) =& \frac{e}{\hbar} \vb{E}_0(t) \cdot \nabla_\mathbf{k} f_\mathbf{k}(t) \nonumber \\
    &+ \Gamma^{in}_\mathbf{k}(t) (1-f_\mathbf{k}(t)) - \Gamma^{out}_\mathbf{k}(t) f_\mathbf{k}(t) \label{eq:boltzmann} ~,
\end{align}
where $\Gamma^{in/out}_\mathbf{k}(t)$ are the electron-phonon scattering matrices according to Fermi's golden rule within a bath assumption in Born-Markov approximation as derived in Ref.~\cite{grumm_theory_2025}. Incorporating normal and umklapp processes, the scattering rates $\Gamma^{in/out}_\mathbf{k}(t)$ taking into account the Pauli principle and include therefore an intrinsic many-body non-linearity. The excitation by a light field $\vb{E}_0$ acting on the electrons is described here as a vectorial quantity and not via the direction-independent intensity as in previous approaches focusing on energy relaxation out of a momentum-isotropic situation \cite{del_fatti_nonequilibrium_2000, pietanza_non-equilibrium_2007, rethfeld_ultrafast_2002, mueller_relaxation_2013, riffe_excitation_2023, ono_ultrafast_2020, sarkar_electronic_2023, seibel_time-resolved_2023, allen_theory_1987, dubi_hot_2019, sivan_theory_2021}. In our approach, describing excitation and orientational relaxation self-consistently, the light field breaks the symmetry of the momentum distribution by the distinct polarization direction of $\vb{E}_0$. Disturbing the momentum direction of the initial isotropic electron gas selectively, this opens the possibility to study orientational relaxation as new relaxation/dephasing channel. 

The Boltzmann equation \eqref{eq:boltzmann} is solved numerically with gold parameters listed in Tab.~2 in Ref.~\cite{grumm_theory_2025}. In the following, we investigate the combined dynamics of electron relaxation in terms of orientational relaxation, thermalization and cooling after optical excitation in a bulk system (cf.~Fig.~\ref{fig:ORT}). We apply a linearly $y$-polarized light field $\vb{E}_0$ represented by a single Gaussian pulse with a duration of $12$~fs (FWHM) and a carrier frequency of $5$~THz. On the expected time scale of the relaxation dynamics described above, this pulse acts as an initial condition which allows to study the intrinsic electron relaxation without the influence of a finite pulse duration or fast oscillating fields but rather the sequence of electron-phonon processes without interference of the initial excitation. 
\\%
In the following, we will study linear and non-linear excitation regimes. Both regimes can be distinguished via the distortion of the electron distribution, i.e.~by the energy the electrons absorbed from the light field $\epsilon_{abs}$ (cf.~Ref.~\cite{grumm_theory_2025}) in comparison with the thermal broadening of the Fermi edge $k_B T_{eq}$: In the regime of non-linear excitations, the field strength is set to $E_0 = 0.7~\mvcm$ ($\epsilon_{abs} \gtrsim k_B T_{eq} $), while for linear excitations $E_0/100$ ($\epsilon_{abs} \ll k_B T_{eq} $) is assumed. In this context, we expect that orientational relaxation without any energy exchange with the phonon system exists across all excitation regimes, while thermalization and cooling as energy relaxation processes occur only in the non-linear regime.

\begin{figure}
    \includegraphics{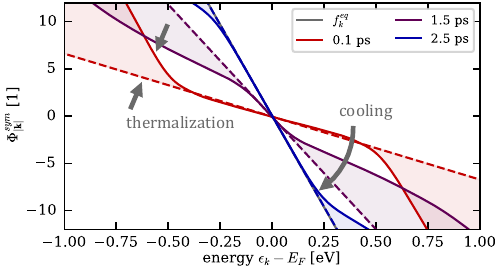} 
    \caption{\label{fig:energyspace_eph_dynamics} The quasi-logarithmic representation of the symmetric distribution $\Phi^{sym}_{|\vb{k}|}$ (solid lines) reveals the thermal (dashed lines) and non-thermal distribution (shaded areas).}
\end{figure}

\textit{Qualitative Discussion.~~}
The light-matter interaction in the Boltzmann equation \eqref{eq:boltzmann} breaks the original radial symmetry of the electron distribution in momentum space, since electrons are initially accelerated in the polarization direction $\vb{e_y}$ of the light field $\vb{E}_0$. Counteracting to this, the electron-phonon interaction does not prefer any direction and thus starts to equilibrate all momentum orientations to a distribution of uniform momentum directions. To analyze this process, we divide the electron distribution into a non-symmetric distribution $f_{k_x,k_y,k_z}^{non-sym}$ depending on the direction of the momentum $\vb{k}=(k_x,k_y,k_z)^T$ and a symmetric distribution $f_{|\vb{k}|}^{sym}(t)$ depending on the momentum absolute $\vert \vb{k} \vert$
\begin{align}
    f_\mathbf{k}(t) =& f_{k_x,k_y,k_z}^{non-sym}(t) + f_{|\vb{k}|}^{sym}(t) ~, \label{eq:sym_nonsym_electron_distribution} \\
    f_{|\vb{k}|}^{sym}(t) =& f_{|\vb{k}|}^{non-th}(t) + f_{|\vb{k}|}^{th}(t) ~. \label{eq:th_nonth_electron_distribution}
\end{align}
Later, to distinguish thermalization and cooling, the symmetric distribution $f_{|\vb{k}|}^{sym}$ is divided into thermal $f_{|\vb{k}|}^{th}$ and non-thermal $f_{|\vb{k}|}^{non-th}$ components \cite{fann_electron_1992, dubi_hot_2019}.
\\%
Thus, as illustrated in Fig.~\ref{fig:ORT}, orientational relaxation applies to the fact that the non-symmetric electron distribution $f_{k_x, k_y, k_z}^{non-sym}$ induced by the light field $\vb{E}_0$ is driven into a symmetric distribution $f_{|\vb{k}|}^{sym}$, which does not prefer a direction in momentum space. Our numerical results indicate that this process of momentum orientational relaxation occurs independently of the optical excitation strength in the linear and non-linear regime and without dissipating energy into the phonon bath. To illustrate this process, Fig.~\ref{fig:kspace_eph_dynamics}a and b depict the temporal occupation dynamics in the $k_x$- (equivalent to $k_z$), and $k_y$-direction (orthogonal and parallel to the incident light polarization) for three different momentum states located below, at and above the Fermi edge. To enhance the visibility of this process, we depict in Fig.~\ref{fig:kspace_eph_dynamics} only the results for the non-linear excitation with strength $E_0$ where the electron gas is strongly perturbed. Nevertheless in the linear regime, orientational relaxation shows the same qualitative signatures:
\begin{figure*}
    \includegraphics{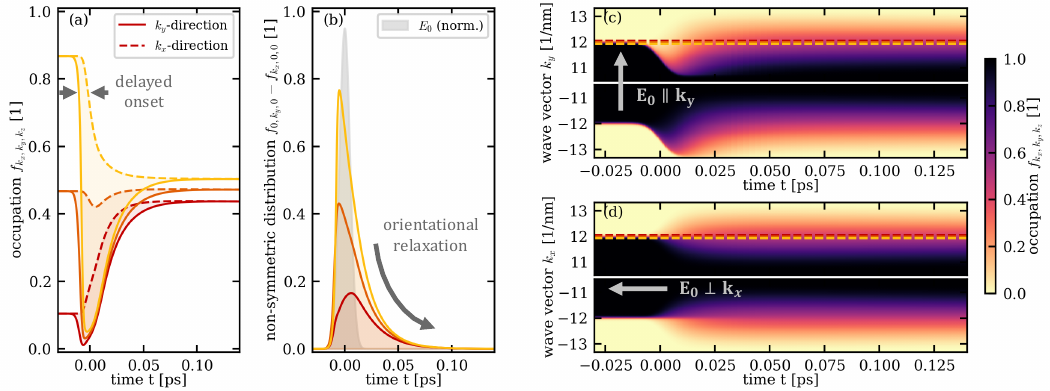} 
    \caption{\label{fig:kspace_eph_dynamics} Dynamics of the electron gas after excitation with a $y$-polarized pulse with strength $E_0$. In (a) the dynamics of the occupations in $k_y$-direction (with $k_x=k_z=0$, solid lines) and in $k_x$-direction (with $k_y=k_z=0$, dashed lines) are shown for different momentum states below, at and above the Fermi edge (dashed lines in c,d correspond to energies of $E_F$ and $E_F \pm 2k_B T_{eq}$, respectively). The decay of the occupation difference between $k_y$- and $k_x$-direction visualizes in (b) the orientational relaxation. The electron distribution in momentum space is depicted along the (c) $k_y$- and (d) $k_x$-axis (the $k_z$-direction is identical with the $k_x$-case) and represents a Fermi-Dirac distribution in equilibrium.}
\end{figure*}
\\%
Out of the initial equilibrium Fermi-Dirac distribution $f_k^{eq}(T_{eq})$ during the first tens of fs after the excitation, the occupations in Fig.~\ref{fig:kspace_eph_dynamics}a clearly show a non-symmetric behavior of the distribution $f_\mathbf{k}$ with respect to $k_x$- and $k_y$-direction. The onset of the dynamics in the $k_x$-direction is temporally delayed compared to the $k_y$-direction which is initially elongated by the light field on the scale of the pulse duration of $12$~fs. After the \textit{orientational relaxation time} $\tau_{or}$ of less than $100$~fs the directional dependence of $f_{k_x,k_y,k_z}$ is temporally synchronized (cf.~Fig.~\ref{fig:kspace_eph_dynamics}b), i.e. equilibrated with respect to the momentum direction and has developed into a symmetric distribution depending only on $|\vb{k}|$, i.e.~on energy $\epsilon_k$. 
Such an orientational relaxation has already been studied in the semi-metal graphene \cite{malic_efficient_2012, mittendorff_anisotropy_2014}, photoexcited semiconductors \cite{binder_greens_1997} or on the basis of a relaxation time approximation in metals \cite{brown_nonradiative_2016, brown_ab_2016}, but as far as we know, has never been discussed quantitatively for noble metals in terms of the full non-linear electron dynamics such as Eq.~\eqref{eq:boltzmann}. 
In Fig.~\ref{fig:kspace_eph_dynamics}c and d, the evolution of the full electron distribution $f_\mathbf{k}$ as function of $k_y$ and $k_x$ is visualized in time and momentum space. Initially, before excitation at $t=0$, c and d show a Fermi-Dirac distribution in equilibrium with the phonon bath and a Fermi level at $k_F = 12.0~\text{nm}^{-1}$ (dark areas describe occupied electron states and bright areas unoccupied states). On the time scale of the excitation pulse in Fig.~\ref{fig:kspace_eph_dynamics}c, the strong displacement of the electron distribution parallel to the polarization of the light field in $k_y$-direction is visible. 
\\%
While in the linear regime the electron gas is back in thermodynamic equilibrium after orientational relaxation and no further relaxation proceeds, in the non-linear regime energy relaxation continues: Energy relaxation includes thermalization as redistribution of the non-thermal electronic energy into thermal one and cooling as energy dissipation to the phonon bath \cite{dubi_hot_2019, mueller_relaxation_2013}. Thus, for non-linear excitations, orientational relaxation constitutes only the first temporal step in a sequence of relaxation events. The processes of thermalization and cooling as a function of energy are well studied in literature \cite{dubi_hot_2019, sivan_theory_2021, del_fatti_nonequilibrium_2000, pietanza_non-equilibrium_2007, rethfeld_ultrafast_2002, mueller_relaxation_2013}. Here, we would like to show that they develop in a consistent fashion out of the same theory as the orientational relaxation, i.e.~by solving Eq.~\eqref{eq:boltzmann} in a momentum-resolved picture:
\\%
After the orientational relaxation, electrons are distributed symmetrically $f_{|\vb{k}|}^{sym}$ but not necessarily thermal (i.e.~non-Fermi-Dirac), cf.~Eq.~\eqref{eq:th_nonth_electron_distribution}, and an energy-dependent representation of the distribution in terms of  $\epsilon_k =\frac{1}{2m}\hbar^2 |\vb{k}|^2$ is permissible. In the next step, the transition from a non-thermal distribution $f_{|\vb{k}|}^{non-th}$ into a thermal Fermi-Dirac distribution $f_{|\vb{k}|}^{th}$ due to electron-phonon scattering proceeds in the \textit{thermalization time} $\tau_{th}^{ep}$. At the same time, the excited hot electron gas dissipates its thermal energy to the phonon system in the \textit{cooling time} $\tau_c$.
\\%
To distinguish between the simultaneously proceeding thermalization and cooling processes, we introduce the quasi-logarithmic representation $\Phi^{sym}_{|\vb{k}|} = - \log{((f^{sym}_{|\vb{k}|})^{-1} -1)}$ of the symmetric distribution as in Ref.~\cite{rethfeld_ultrafast_2002}. This function connects a thermal Fermi-Dirac distribution linearly to the electron energy $\epsilon_k$ with slope proportional to the inverse electron temperature $T_e^{-1}$. The solid lines in Fig.~\ref{fig:energyspace_eph_dynamics}a illustrate our numerical result for different times after excitation with the corresponding thermal distribution $f_{|\vb{k}|}^{th}$ as fitted Fermi-Dirac distribution (dashed lines). Their increasing slope over time consequently describes the cooling of the electron gas, whereas the decreasing deviation of the symmetric distribution from the thermal distribution describes thermalization (shaded areas). The complete relaxation of the electron gas from a hot distribution shortly after the excitation back to the equilibrium proceeds on times longer than $2.5$~ps. 

\begin{figure}
    \centering
   \includegraphics{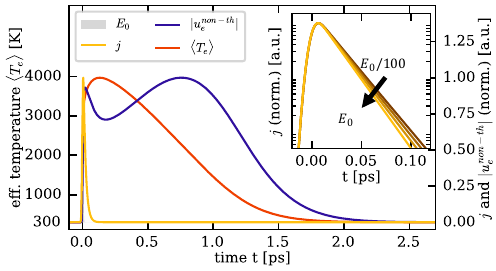}
    \caption{\label{fig:relaxation_times} The current density $j$ decays exponentially within the orientational relaxation time $\tau_{or}$. On longer time scales, the non-thermal energy density $|u^{non-th}|$ and effective temperature $\langle T_e \rangle$ track thermalization and cooling. The inset shows the decay of the current density for different excitation strengths from $E_0/100$ (dark) to $E_0$ (bright).}
\end{figure}

For a more quantitative discussion of the time scales of orientational relaxation, thermalization and cooling, we connect the microscopic distributions in Eq.~(\ref{eq:sym_nonsym_electron_distribution},\ref{eq:th_nonth_electron_distribution}) to observables:

\textit{a) Orientational relaxation.~~}
To discuss the orientational relaxation time, we study the macroscopic current density 
\begin{align}
    \vb{j}(t) = -\frac{2e}{V} \sum_\mathbf{k} \vb{v_k} f_{k_x, k_y, k_z}^{non-sym}(t) \label{eq:current_density}
\end{align}
defined via the electron velocity $\vb{v_k} = \frac{1}{\hbar} \frac{\partial \epsilon_k}{\partial \mathbf{k}}$. From symmetry considerations follows that the current density is carried exclusively by the non-symmetric distribution $f_{k_x,k_y,k_z}^{non-sym}$ \cite{hess_maxwell-bloch_1996-1}. This allows us to determine the orientational relaxation time directly from the temporal decay of the current density, Eq.~\eqref{eq:current_density}, shown in Fig.~\ref{fig:relaxation_times} (yellow line and inset). In the regime of linear excitations with $E_0/100$ (darkest line in the inset), the decay time corresponds to an orientational relaxation time of $\tau_{or}=19.7$~fs. In contrast, for non-linear excitations with field strength $E_0$ (brightest line in the inset) the orientational relaxation time decreases to $\tau_{or}=17.1$~fs. This implies a reduction of the orientational relaxation time for stronger excitations, if the amount of absorbed energy $\epsilon_{abs}$ is at least in the same order of magnitude as the thermal broadening $k_B T_{eq}$.
\\%
The current density $\vb{j}$ acts as source term in Maxwell's equations and is consequently essential for the optical response of the noble metal. As we report in Ref.~\cite{grumm_theory_2025}, the orientational relaxation rate $\tau_{or}^{-1}$ is in the linear regime identified with the damping rate $\gamma_D$, introduced phenomenologically in the classical Drude model \cite{drude_zur_1900, maier_plasmonics_2007}. The Drude model assumes a field strength-independent damping in agreement with our observation for the orientational relaxation rate in the linear regime. 
Experimental studies on the optical dielectric function of gold reveal damping times between $9$~fs and $26$~fs for excitations below the optical band gap \cite{johnson_optical_1972, ordal_optical_1987, babar_optical_2015, magnozzi_plasmonics_2019, olmon_optical_2012}. \textit{Ab initio} studies based on a relaxation time approximation predict values between $24.0$~fs and $26.3$~fs \cite{brown_nonradiative_2016, mustafa_ab_2016} in agreement with our results in the linear regime. 

\textit{b) Energy Relaxation.~~}
In the regime of linear excitations, the energy transfer by the light field to the electron gas is negligible ($\epsilon_{abs} \ll k_B T_{eq}$) and the definition of a thermalization and cooling time is unnecessary. In the non-linear regime, energy relaxation occurs for non-thermal and thermal electrons via an energetic redistribution and dissipation.
\\%
To access the dynamics of non-thermal electrons, we define the non-thermal energy density \cite{rethfeld_ultrafast_2002} 
\begin{align}
    u_e^{non-th}&(t) = \frac{1}{\pi^2} \int_0^\infty \mathrm{d}k ~k^2 \epsilon_k (f_{\vert \vb{k} \vert}^{sym}(t) - f_{\vert \vb{k} \vert}^{th}(t)) ~.\label{eq:non_thermal_energy_density} 
\end{align}
The thermal distribution is addressed by a Fermi-Dirac function
\begin{align}
    f_{|\vb{k}|}^{th}(t) = \frac{1}{\exp{\frac{\epsilon_k-\langle \mu(t) \rangle}{k_B \langle T_e(t) \rangle}}+1} \label{eq:fermi_dirac}
\end{align}
where the effective temperature $\langle T_e \rangle$ and the chemical potential $\langle \mu \rangle$ are defined by the slope of the function $\Phi^{sym}_{|\vb{k}|}$ at the Fermi level (cf.~Fig.~\ref{fig:energyspace_eph_dynamics}). It is important to recognize that this effective temperature is not a function of state of the total electron gas but is defined exclusively for the thermal distribution and is only a measure for the thermal energy.
\\%
As shown in Fig.~\ref{fig:relaxation_times}, the thermalization, following on orientational relaxation, analyzed in terms of the non-thermal energy density reveals a more complex dynamics than a simple exponential decay. Here, a definition of a simple relaxation time is not possible but non-thermal electrons participate on a time span of $\tau_{th}^{ep}\approx 2.0$~ps in the relaxation dynamics. The decay and subsequent re-increase of the non-thermal energy $|u^{non-th}|$ between $0.1$~ps and $0.8$~ps after excitation suggests that electron-phonon scattering first leads to a state more distant from equilibrium before the final thermalization proceeds. Similar observations have already been made for thermalization by electron-electron scattering \cite{seibel_time-resolved_2023}.
The thermalization, described by our results exclusively in terms of electron-phonon interaction, is in agreement with observations in Ref.~\cite{rethfeld_ultrafast_2002, mueller_relaxation_2013}. Nevertheless, for a comprehensive description of thermalization, which is not the scope of this paper, electron-electron interaction has to be considered.
\\%
Simultaneous to thermalization, the electron gas dissipates thermal energy to the phonon bath in a cooling time $\tau_c = 0.8$~ps (red line in Fig.~\ref{fig:relaxation_times}). Since heating and cooling of the electron gas are in a microscopic picture intrinsically non-linear processes, the exact times depend on the strength of the light field and increase with increasing excitation strength. Similar excitation dependent cooling times in gold are provided by versatile experimental and theoretical approaches between $0.8$~ps and $1.2$~ps \cite{fann_direct_1992, sun_femtosecond-tunable_1994, sun_femtosecond_1993, groeneveld_femtosecond_1995, del_fatti_nonequilibrium_2000, giri_experimental_2015, brown_experimental_2017, zhang_plasmon-driven_2019, suemoto_relaxation_2021}. 
\\%
Finally, the electron gas reaches the thermodynamic equilibrium after about $2.5$~ps, when the non-thermal distribution is decayed completely and the temperature $\langle T_e\rangle = T_e = T_{eq}$ becomes a state function of the full electron gas \cite{puglisi_temperature_2017, dubi_hot_2019}. 

\textit{Conclusion.~~}
We studied a momentum-resolved Boltzmann equation for bulk gold to fill the gap between ultrafast optical excitation providing a momentum-polarized electron gas to its energy-resolved thermalization and cooling. During orientational relaxation, the non-symmetric electron distribution dominates for short times less than $100$~fs directly after the excitation. Afterwards for non-linear excitations during thermalization, the symmetric electron distribution develops out of a non-thermal into a thermal distribution and simultaneously relaxes back to the equilibrium by dissipating energy to the phonons in about $2.5$~ps. Since all three relaxation processes depend on the strength of the excitation, a model for the non-linear optical response of noble metals must include many-body interactions such as Pauli blocking, shown here exemplarily for electron-phonon interaction.
In future work, the impact of electron-electron interaction on the orientational relaxation has to be examined to achieve a complete picture of ultrafast electron relaxation in noble metals. In addition, as shown in a first example in Ref.~\cite{grumm_theory_2025}, a self-consistent treatment of Maxwell's equations has to be added to describe the dephasing of plasmonic excitations in spatially finite nanostructures.
\begin{acknowledgments}
We acknowledge fruitful discussions with Robert Salzwedel, Malte Selig, Lara Greten, Joris Strum, Robert Lemke (TU Berlin), Francesca Calegari (DESY Hamburg), and Holger Lange (Uni Potsdam).
\end{acknowledgments}
\bibliography{references}
\end{document}